\documentclass[aps,preprint]{revtex4}%
\usepackage{amsfonts}
\usepackage{amsmath}
\usepackage{amssymb}
\usepackage{graphicx}%
\begin{document}
\title[Wigner distribution]{Information-theoretic significance of the Wigner distribution}
\author{B. Roy Frieden}
\affiliation{College of Optical Science, Univ. of Arizona, Tucson, AZ 85721}
\author{Bernard H. Soffer}
\affiliation{Dept. of Electrical Engineering, UCLA, Los Angeles, CA 90024}

\begin{abstract}
A coarse grained Wigner distribution $p_{W}(x,u)$ obeying positivity derives
out of information-theoretic considerations. \ Let $p(x,u)$ be the unknown
joint PDF\ (probability density function) on position- and momentum
fluctuations $x,u$ for a pure state particle. \ Suppose that the phase part
$Psi(x,z)$ of its\ Fourier transform $F.T.[p(x,u)]=|Z(x,z)|\exp[iPsi(x,z)]$ is
constructed as a hologram. (Such a hologram is often used in heterodyne
interferometry.)\ Consider a particle randomly illuminating this phase
hologram. \ Let its two position coordinates be measured. \ Require that the
measurements contain an extreme amount of Fisher information about true
position, through variation of the phase function $Psi(x,z).$\ The extremum
solution gives an output PDF $p(x,u)$ that is the convolution of the Wigner
$p_{W}(x,u)$ with an instrument function\ defining uncertainty in either
position $x$ or momentum $u$. \ The convolution arises naturally out of the
approach, and is one-dimensional, in comparison with the two-dimensional
convolutions usually proposed for coarse graining purposes. \ The output obeys
positivity, as required of a PDF, if the one-dimensional instrument function
is sufficiently wide. \ The result holds for a large class of systems: those
whose amplitudes $a(x)$\ are the same at their boundaries (Examples: states
$a(x)$\ with positive parity; with periodic boundary conditions; free particle
trapped in a box).

\end{abstract}
\maketitle
\tableofcontents

\section{\ \bigskip Introduction}

\emph{Note}:\ To avoid confusion, the word "phase" below is reserved to
describe only the phase part of a complex amplitude. \ "Phase"\ is never used
to describe "phase space" of statistical mechanics, i.e. joint position and
momentum values $(x,\mu).$ \ These are always demarked as "position-momentum"
space or $(x,\mu)$\ space.

Consider a single, mass particle moving in one dimension and in a pure state
$\psi(x),$ where the random variable $x$ defines an \emph{intrinsic
fluctuation}, i.e. one that would exist even in the presence of a perfect
(noise free) detector. \ The state $\psi(x)$ can be\ defined e.g. by the
nonrelativistic Schrodinger wave equation. \ By Fourier transform of $\psi
(x)$, this also gives the\ particle's probability amplitude $\varphi(\mu)$ on
intrinsic momentum $\mu$.\ \ Let these two amplitude laws be known. \ (Note
that all functions in this analysis depend as well upon the time; for brevity,
this is suppressed from the notation. )

Note that $\psi(x)$ and $\varphi(\mu)$\ are single-variable, \emph{marginal}
probability amplitudes, leaving open the question of the \emph{joint}
dependence of the joint fluctuations $(x,\mu).$ Quantum mechanics, regarded as
a statistical theory, is not fully consistent probabilistically, since it does
not make use of, or define, joint or conditional probabilities such as
$p(x,\mu)$, $p(x|\mu)$, $p(\mu|x)$, etc. \ Here we consider the question of
what the joint PDF (probability density function) $p(x,\mu)$ should be. \ How
should it relate to $\psi(x)?$ \ Is there a universal PDF $p(x,\mu)$, i.e. a
unique function of $\psi(x),$\ or should the function depend upon the
particulars of the \emph{given} measurement scenario?

\ Wigner [1] proposed the well-known joint PDF%

\begin{equation}
p_{W}(x,\mu)=\frac{1}{2\pi}\int dz~e^{-~iz\mu}\psi^{\ast}(x-\hbar
z/2)\psi(x+\hbar z/2)
\end{equation}
for constructing a measure of the joint fluctuations in position-momentum
space. Here $\hbar$ is Planck's constant/$2\pi$ and $i=\sqrt{-1}.$ Conversely,
given a $p_{W}(x,\mu)$ obeying (1) the wave function $\psi(x)$ may be
reconstructed to within an irrelevant constant phase value. \ Hence (1) is
often considered to be a generally complex quantum formulation that is
equivalent to Schrodinger's. \ It is useful for visualizing the joint
evolution of the position-momentum values. Result (1) has also been shown to
follow from various operational viewpoints [2] (see as well the extensive
bibliography and background for the problem given in [2]).

\ Unfortunately, for a \emph{general} state function $\psi(x)$ Eq. (1) is
known to incur negative values and, hence, cannot represent a well defined
probability law. \ \ (The only case that does not incur negatives is the
normal case, including squeezed or chirped versions.)\ This is also consistent
with limitations set by the Heisenberg uncertainty principle, according to
which precise joint values $(X,M)$ of position and momentum \emph{do not
exis}t on the quantum level [3],[4]. \ 

These limitations are taken to imply that, given a marginal amplitude
$\psi(x),$ there is \emph{no singl}e joint PDF $p(x,\mu)$ that is generally
well defined. \ Again, this is for \emph{intrinsic} fluctuations $(x,\mu
).$\ On the other hand, a \emph{real} measurement scenario, whether
experimental or gedanken, is guaranteed to \emph{obey} a well-defined PDF on
its \emph{total} fluctuations (including noise of detection).\ Therefore, from
here on, \emph{by }$(x,\mu)$\emph{\ we mean total fluctuations in position and
momentum.}\ \ 

\ In summary, our view is that a variety of physically meaningful PDFs
$p(x,\mu)$ exist, where each is defined for the \emph{particular} measurement
scenario out of which it derives. \ In general, such a law depends upon the
physics of the measurement scenario, both through the state $\psi(x)$
\emph{and} through properties of the detector or other influences on the
measurement. \ \ This general view was previously taken [5] as well (see note
at ref.). \ Consequently the PDF $p(x,\mu)$ that we obtain below is limited in
validity to a particular gedanken measurement scenario, namely that of
particle location in a hologram. \ An immediate benefit is that, in describing
a real measurement, the fluctuations $(x,\mu)$ must now describe those in the
\emph{total experimental }measurement, including possible noise of detection.
\ Thus, a theory of measurement emerges.

An ad hoc supplement to (1) that forces positivity is to mathematically
convolve (1) with a chosen kernel function [2]. \ This could have a physical
origin in coarse graining [2] the space $(x,\mu)$. \ The resulting measure is
then taken to describe the joint probability law of the coarse grained space.
\ \ Coarse graining has so far been proposed, by convolution of (1) with a
suitably broad kernel function in $x$ and $\mu,$ for example a Gaussian.\ This
results from the well-known result that the convolution of two Wigner
distribution functions obeys positivity [2]. \ The minimum amount of coarse
graining that suffices to give positive Wigner values has been established [6]
as that obeying the Heisenberg uncertainty principle.\ \ 

This convolution step is usually implemented as an ad hoc mathematical add-on
to $p_{W}(x,\mu).$\ \ By comparison, our view is, as above, that a valid
convolution step\emph{\ should be a consequence of} the physics of the
particular measurement problem.\ \ Indeed, by our approach it will be found to
follow as a consequence of determining a particle position in a phase
hologram. \ \ In its emphasis upon measurement, the approach is reminiscent of
a previous analysis [5], which showed that a positive-constrained, Wigner-like
PDF results from considering a scattering experiment in the Born
approximation. \ \ Such Wigner-like PDFs have likewise followed as the outputs
of optical heterodyne imaging experiments [7]-[11] wherein phase object
profiles are estimated.

Such a convolution is equivalent to blurring at the microlevel of
\emph{points} $(x,\mu).$ \ This blurring will have an important consequence to
the point-level $J$ of \emph{information}. \ This is that the microlevel
information level $J=0$ (see Sec. 2.4).

\ We next show that Wigner's function (1) follows from this overall viewpoint.
\ In particular this will be out of the gedanken measurement of a particular
scenario. \ This is of the position and momentum of a particle irradiating a
holographic object. \ The scenario is suggested by past successful Wigner-like
answers for optical heterodyning [7]-[11] methods, which likewise serve to
determine phase objects. \ The analytical approach to be used is that of
Extreme physical information (EPI) [12]-[15]. \ (Note that EPI avoids the use
of standard operator quantum mechanics.) \ The approach is handy in being
basically statistical in nature, thereby enabling both quantum and classical
statistical effects to be derived. \ Indeed, regarding the requisite
convolution (above), [5] "The idea that in any realistic measurement a
detector and a filtering device [as here] are required is not really quantum
mechanical in nature." \ \ \ 

The EPI approach is well-suited to the problem, since (a) it has a strong
track record of deriving probability amplitude laws [12]-[15]; and (b) has
derived both quantum and classical PDFs. \ The result will be the convolution
of the quantum Wigner law $p_{W}(x,\mu)$ with a classical noise distribution
on either momentum or position. \ \ We emphasize that this is a \emph{one}%
-dimensional convolution in place of the usual two-dimensional one mentioned
above. \ Also, the statistical nature of the approach will allow the
convolution kernel to be interpreted as straightforwardly a PDF on noise of detection.

\section{EPI Approach}

\ EPI is a general approach for calculating amplitude laws, PDFs, and
input-output laws for the fluctuations of unknown systems. \ The approach
centers on the flow of information that occurs during the measurement of a
required parameter by an observer. \ EPI is briefly defined in the
introductory paper [12], and fully developed in the books [13],[14].

\subsection{ Extremum condition\ }

The EPI approach requires solving an extremum problem
\begin{equation}
I-J=extrem.
\end{equation}
for the system amplitude or probability law. \ In general $I$ is the Fisher
information in the data and $J$ is that in the source. \ Eq. (2) states that
the loss of information from source to data is an extreme value (usually a
minimum). This condition is the central \emph{ansatz} of EPI, and has been
abundantly verified [13],[14] by application. \ The \emph{ansatz} is obeyed
rigorously [12]-[14]\ in the presence of a unitary transformation.
\ Accordingly, an obvious transformation of this type will be utilized.
\ Also, it will be shown below that effectively $J=0$ in this problem, so that
the EPI principle Eq. (2) simplifies here to $I=extrem$.

\subsection{Rotation space}

\ \ All quantum EPI calculations start with a rotation of either coordinates
or amplitude functions. \ Such a rotation is demanded by the length-preserving
nature of Fisher information under unitary transformation [13],[14]. \ Note,
e.g., that information quantity (11) is a sum (integral) of squares and,
hence, invariant under such transformation. \ As in (11), which is an integral
over the space of the \emph{amplitude} $\Psi,$ the invariance is specifically
with respect to \emph{amplitude} (not PDF) laws. \ Past examples of such
rotations are from four-position space into four-momentum space in deriving
the Klein-Gordon, Dirac, and Wheeler-DeWitt equations of quantum mechanics
[12]-[14]; rotation by a complex angle in deriving the Lorentz transformation
of special relativity [14]; and rotation by the Weinberg angle in Higgs mass
theory [14].

\ In an unknown scenario, the user has to use \emph{physical intuition} in
choosing the appropriate rotation.\ \ However, as an aid, EPI\ is
\emph{exhaustive} under such rotations [14] :\ \ In practice every
well-defined rotation leads to a new \emph{physical} solution for the
amplitude function $\psi.$\ What, then, should be rotated here?

The physical intuition here is that \emph{complex} object distributions
$Z(x,z)$ tend to be well approximated by their \emph{phase parts }$\Psi(x,z).
$ In fact, it was shown by Kermisch [16] for a class of holograms that the
information about photon locations in $Z(x,z)$ is carried by about $78\%$ of
the photons that form $\Psi(x,z).$ \ This is one reason why phase-only
holograms are practical as information storage devices. \ Also, phase
distributions are noted for having high local gradients [17], and Fisher
information - a local measure of information - is notably sensitive to such
gradients. \ \ This further agrees with the need for using the channel
capacity form of $I$, i.e. its \emph{maximized} form (see above).

The preceding two paragraphs suggest that the rotation for the problem be a
F.T. operation on position-momentum space [2],%

\begin{equation}
Z(x,z)\equiv\int d\mu~p(x,\mu)e^{iz\mu}\equiv|Z(x,z)|\exp[i\Psi(x,z)].
\end{equation}
Function $Z(x,z)$ will be our hologram, as shown below, with $\Psi(x,z)$ its
phase part. \ Such a two-dimensional phase hologram $\Psi(x,z)$\ can in
principle be formed optically [18], digitally [19] by generation of computer
holograms, or by other means.\ \emph{The EPI gedanken measurement for this
problem will accordingly be that of the theoretical position }$X,Z$\emph{\ of
a\ particle in the phase hologram }$\Psi(x,z).$ \ This will allow $\Psi(x,z)$
to be reconstructed from the principle. \ \ However, we emphasize that this is
a gedanken measurement: \ neither the hologram or the measurement is actually
implemented. \ Hence, our results express a contingency:\ \ \emph{If} such a
hologram were formed, and measured, then the EPI principle implies that the
unknown joint PDF $p(x,\mu)$ would be the convolution mentioned above.\ \ \ In
this way, \emph{the PDF }$p(x,\mu)$\emph{\ is seen to represent an ideal state
of information}, as occurs in deriving other laws of physics via EPI [12]-[14].

The problem of reconstructing a phase hologram has a long history,
particularly by the use of heterodyne interferometry [7]-[11]. \ Here the
latter is replaced by the use of an EPI gedanken measurement, and accompanying
use of principle (2).

By the completeness of the Fourier description (3), an EPI\ problem of
estimating $p(x,\mu)$ is thereby replaced with the problem of estimating the
function $Z(x,z).$ \ Once the latter is known, (3) shows that $p(x,\mu)$ may
be computed as%

\begin{equation}
p(x,\mu)=F.T.(Z)\equiv\frac{1}{2\pi}\int dzZ(x,z)e^{-iz\mu}~\equiv~\ \frac
{1}{2\pi}\int dz|Z(x,z)|\exp[i\Psi(x,z)]e^{-iz\mu}.
\end{equation}

However, there is a drawback to this mathematical approach. \ As mentioned
above, Fisher information is an invariant $L^{2}$ length in amplitude space
and not PDF space.\ Hence, the rotation (3) in PDF space does not strictly
comply with EPI. \ This is confirmed by the fact that, in depending upon the
PDF $p(x,\mu),$ it ignores phase information in the amplitude function whose
square is $p(x,\mu).$\ Hence the answer we get must be approximate, lying
somewhere between quantum and classical physics.

\subsection{Holographic aspect of rotation}

We next show that $Z(x,z)$ and its phase part $\Psi(x,z)$\ can be repesented
as holograms in two-dimensional position space. \ This requires showing
that\ coordinate $z$ is proportional to a position. \ As usual,
position-momentum space is subdivided into elemental cells of sides $\Delta
l,\Delta\mu,$ with areas $\Delta l\Delta\mu=\hbar/2$ by the Heisenberg
principle. \ Hence, a position determination $l_{z}=n_{z}\Delta l,$ for an
appropriate integer $n_{z}.$ Combining the last two relations gives
$l_{z}=n_{z}\hbar/(2\Delta\mu).$ Also, in the exponent of \ (4), $z\mu$ must
be unitless, so that we may express
\begin{equation}
z=n_{z}/\Delta\mu.
\end{equation}
\ Combining the last two relations gives\
\begin{equation}
l_{z}=\ (\hbar/2)z\text{ and }dl_{z}=\ (\hbar/2)dz
\end{equation}
as its differential. \ Hence $z\propto l_{z},$ or, the transform space
coordinate $z$ is effectively back in position space defining a corresponding
position coordinate $(\hbar/2)z$. \ Since coordinate $x$ is likewise a
position, effectively function $Z(x,z)$ and $\Psi(x,z)$\ lie entirely in
position space. \ This facilitates the required position measurement ($X,Z$),
which will now be of the rectangular position in a dimensional hologram (see next).

\subsection{ Reconstruction step\ }

In the reconstruction step, the phase hologram $\Psi(x,z)$ be illuminated with
a uniform plane wave called a \emph{reference beam}. \ Here either material
particles or photons may be used. \ \ To be definite, we choose material particles.

The aim of the gedanken experiment is to measure the ideal joint position
values $(X,Z)$ of a randomly chosen particle as it passes through the phase
hologram. \ In practice, the measurement is imperfect, with respective error
fluctuations $(x,z).$\ \ The question that EPI will seek to answer is, What
phase profile $\Psi(x,z)$ extremizes the acquired information $I$ about
position $(X,Z)$\ in the data? \ Once this is known, its use in (4) gives the
required PDF $p(x,\mu)$ on position and momentum for the illuminating particle.

As mentioned above, such phase profiles $\Psi(x,z)$ have previously been found
using heterodyne approaches [7]-[11], and these do tend to follow Wigner-type
distributions, as required. \ 

Note from (3) that $Z(x,z)$ is generally complex so that \ $\Psi(x,z)$ is not
necessarily real. \ This will not matter to the calculation.\ \ The phase law
$\Psi(x,z)$ has a corresponding intensity profile%
\begin{equation}
P(x,z)\equiv\Psi^{\ast}(x,z)\Psi(x,z)
\end{equation}
where $^{\ast\text{ }}$denotes the complex conjugate. \ This is also the PDF
on particle positions in the phase hologram. \ \ 

\subsection{Parameters to be measured}

To review, the unknown parameters that are to be gedanken-measured are the $X
$-position and effective $Z$-position (as above) of a randomly selected
particle that passes through the phase hologram $\Psi(x,z).$ \ These are
measured with respective errors $x$ and $z$, using an instrument that
generally suffers from noise. \ The total measurement errors \ $(x,z)$\ are
therefore inclusive of both this noise and the holographic object.\ \ We next
seek the PDF on these total measurement errors using EPI.

\subsection{Source information $J$}

The general flow of information in an EPI measurement procedure is from the
information source to the measurement space,%

\begin{equation}
J\rightarrow I.
\end{equation}
Here, by definition the information source is at the \emph{point} level
$(x,z)$\ of the joint position fluctuations. \ However, there is effectively
no given information at that level since data averaging (coarse graining) will
be taken in \ $(x,\mu)$ space, as discussed above,\ and this causes effective
data averaging in $(x,z)$ space as well, via the Fourier transform operation
(4). That is, coarse graining for the original particle scenario translates
into coarse graining in the hologram scenario as well. \ Hence, in this
calculation \ %

\begin{equation}
J=0.
\end{equation}
There is effectively no information on the microscale of this gedanken
experiment. Note that this coarse graining will be a result of the
calculation, not an assumption. \ That is, the EPI\ solution to the problem
will be self-consistently smeared out sufficiently in momentum or in position
\emph{to} \emph{have} this property.

In general, use of $J=0$ indicates an EPI calculation of lowest precision,
level (c) [13],[14]. \ Results must be regarded as approximate or contingent
(as with the Wigner answer (1), which is contingent upon coarse graining or a
Gaussian $\psi(x)$). \ This is the second approximation that is made in the
overall approach. \ The other was the rotation in PDF or energy space rather
than in amplitude space (see above).

A check on the assumption $J=0$ will be the calculated $I$ at solution.
\ Since $I=\kappa J$ according to EPI theory, then the solution should obey
$I=0$ as well. \ This is verified in the Appendix for a class of state
functions $\psi(x).$

\subsection{Data information $I$}

\ With $J=0$, the entire calculation Eq. (2) hinges on the information
functional $I$. \ \ \ The positional errors of the problem are the positions
$x$ and $(\hbar/2)z$ (see (6)). \ \ We choose to Wick-rotate the latter into
an imaginary coordinate $i(\hbar/2)z$ (see above remarks about rotation and
EPI). The Fisher coordinates of the problem are accordingly%
\begin{equation}
(x_{1},x_{2})=(x,i(\hbar/2)z).
\end{equation}
Note that this rotation is arbitrary, and represents prior knowledge on the
part of the observer. \ As usual in EPI\ problems, it will be justified on the
grounds that the solution is reasonable and gives new insight into the
problem. \ The same Wick rotation is of course commonly used to represent the
time coordinate in relativistic effects [20].

With these as Fisher coordinates, quantity $I$\ has the significance of being
the Fisher information in an attempt at measuring the ideal coordinates
$(X,i(\hbar/2)Z)$ of a randomly selected particle in the phase hologram. \ \ \ 

The differentials of the Fisher coordinates are $dx_{1}=dx,$ $dx_{2}\equiv
dl_{z}=i(\hbar/2)dz$ by (6), so that\ $dx_{1}|dx_{2}|=$\ $(\hbar
/2)dxdz.$\ Then the Fisher channel capacity information is [12]-[14]%

\begin{equation}
I=\frac{8}{\hbar}\int\int dxdz\left[  \left(  \frac{\partial\Psi^{\ast}%
}{\partial x}\right)  \left(  \frac{\partial\Psi}{\partial x}\right)  -\left(
\frac{2}{\hbar}\right)  ^{2}\left(  \frac{\partial\Psi^{\ast}}{\partial
z}\right)  \left(  \frac{\partial\Psi}{\partial z}\right)  \right]
,~~~P=\Psi^{\ast}\Psi.
\end{equation}
The minus sign arises out of squaring the imaginary $i$ in $dx_{2}$ ([14],
App. C). The phase profile $\Psi(x,z)$ that attains extreme information in the
data therefore has this level of information $I$.

\section{\bigskip EPI Implementation}

The EPI\ extremization principle is $I-J=extrem.$ \ Then by (9) and (11), the
principle is%

\begin{equation}
\int\int dxdz\left[  \left(  \frac{\partial\Psi^{\ast}}{\partial x}\right)
\left(  \frac{\partial\Psi}{\partial x}\right)  -\left(  \frac{2}{\hbar
}\right)  ^{2}\left(  \frac{\partial\Psi^{\ast}}{\partial z}\right)  \left(
\frac{\partial\Psi}{\partial z}\right)  \right]  =extrem.,~~\Psi=\Psi(x,z).
\end{equation}
We ignored an irrelevent multiplicative constant. \ The rest is algebra.

\subsection{Forming Lagrangian \ }

The Lagangian of the problem is directly the integrand of (12),%

\begin{equation}%
\mathcal{L}%
=\left(  \frac{\partial\Psi^{\ast}}{\partial x}\right)  \left(  \frac
{\partial\Psi}{\partial x}\right)  -\left(  \frac{2}{\hbar}\right)
^{2}\left(  \frac{\partial\Psi^{\ast}}{\partial z}\right)  \left(
\frac{\partial\Psi}{\partial z}\right)  .
\end{equation}

\subsection{Semi-classical Solution}

The general Euler-Lagrange solution obeys%

\[
\frac{\partial}{\partial x}\frac{\partial%
\mathcal{L}%
}{\partial(\partial\Psi^{\ast}/\partial x)}+\frac{\partial}{\partial z}%
\frac{\partial%
\mathcal{L}%
}{\partial(\partial\Psi^{\ast}/\partial z)}=\frac{\partial%
\mathcal{L}%
}{\partial\Psi^{\ast}}.
\]
Hence by (12) the Euler-Lagrange solution to this problem obeys%

\begin{equation}
\frac{\partial^{2}\Psi}{\partial x^{2}}-\left(  \frac{2}{\hbar}\right)
^{2}\frac{\partial^{2}\Psi}{\partial z^{2}}=0.
\end{equation}

This is a wave equation for two traveling waves of "velocities" $\pm\hbar/2 $
with $z$ standing in for the "time." \ Thus, as first shown by Tatarskii [21],
the solution is the sum of these waves%

\begin{equation}
\Psi=F_{1}(x+\frac{\hbar}{2}z)~+F_{2}(x-\frac{\hbar}{2}z),~\ \ F_{1}%
,F_{2}~\text{arbitrary.}%
\end{equation}

Then by (3),%

\begin{align}
Z  &  =|Z|~\exp[iF_{1}(x+\frac{\hbar}{2}z)]\exp[iF_{2}(x-\frac{\hbar}{2}z)]\\
&  \equiv|Z|~f_{1}(x+\frac{\hbar}{2}z)f_{2}(x-\frac{\hbar}{2}z)\nonumber
\end{align}
after defining%

\begin{equation}
f_{1}(x)=e^{iF_{1}(x)},~~f_{2}(x)=e^{iF_{2}(x)}.
\end{equation}
Then by (4) and (16),%

\begin{equation}
p(x,\mu)\equiv F.T.(Z)=F.T.\left[  |Z|~f_{1}(x+\frac{\hbar}{2}z)f_{2}%
(x-\frac{\hbar}{2}z)\right]  .
\end{equation}
Then by the theorem for the F.T. of a product,%

\begin{equation}
p(x,\mu)=g(\mu|x)\otimes p_{W}(x,\mu)\equiv\int d\mu^{\prime}g(\mu-\mu
^{\prime}|x)p_{W}(x,\mu^{\prime}),
\end{equation}
where $\otimes$ denotes a one-dimensional convolution, the kernel function
$g(\mu|x)$ obeys%

\begin{equation}
g(\mu|x)\equiv F.T.(|Z|)\equiv\frac{1}{2\pi}\int dz~e^{-iz\mu}|Z(x,z)|,
\end{equation}
and $p_{W}(x,\mu)$ obeys%

\begin{equation}
p_{W}(x,\mu)=F.T.[f_{1}(x+\frac{\hbar}{2}z)f_{2}(x-\frac{\hbar}{2}z)]=\frac
{1}{2\pi}\int dz~e^{-iz\mu}f_{1}(x+\frac{\hbar}{2}z)~f_{2}(x-\frac{\hbar}%
{2}z).
\end{equation}
The probabilistic notation $|$ in $g(\mu|x)$ denotes the conditional "if",
that is, $g(\mu|x)$ is the\ probability density on a random value of momentum
$\mu$ in the presence of ("if") a fixed value of $x$.

Finally, as noted [5], if we regard $f_{1}(x)$ and$~f_{2}(x)$ as square
integrable, and $p_{W}(x,\mu)$ as real and normalized, then Eq. (21) takes the
particular Wigner form Eq. (1)%

\begin{equation}
p_{W}(x,\mu)=\frac{1}{2\pi}\int dz~e^{-~iz\mu}\psi^{\ast}(x-\hbar
z/2)\psi(x+\hbar z/2)
\end{equation}
for the appropriate choice of functions%

\begin{equation}
f_{1}(x)=\psi(x),~~~f_{2}(x)=\psi^{\ast}(x).
\end{equation}

Eqs. (19) and (22) are the main results of the paper. \ These show that the
EPI solution for the PDF of the measurement problem is the Wigner function
convolved with a kernel function $g(\mu|x)$ along $\mu.$\ One can regard the
Wigner function as the quantum part of the answer, with the convolving kernel
function a classical part. \ The convolution is only along the momentum
coordinate $\mu.$\ Statistically, such a convolution denotes the presence of
added classical noise of \emph{detection} [22] of momentum. \ Hence, the
solution pictures a measurement scenario where the detection is generally
imperfect, suffering from noise in the momentum reading. \ The noise is
classically characterized by a PDF $g(\mu|x)$. \ As above, this means the
probability of a random variable value $\mu,$ in the presence of a general but
fixed value of $x.$ \ \ Hence the noise in $\mu$ is "signal dependent," if we
regard fluctuation value $x$ as being the signal. \ \ This noise effectively
induces coarse graining of momentum space in this problem.\ \ The kernel
function $g(\mu|x)$, as a measure of detection noise, is often called the
\emph{instrument function} of the measurement experiment.

This result distinguishes the solution from most past approaches [2] to a
practical Wigner function in two ways: Past approaches postulate (not derive,
as here) convolution with a kernel function, and in \emph{two-dimensions
}rather than the one dimension here. \ Two-dimensional convolution is
well-known to achieve positivity if the kernel function is chosen to be
another Wigner distribution (in particular, a Gaussian of sufficient width). \ 

Although of secondary importance, the estimated \emph{amplitude} law
$\Psi(x,z)$ obeyed by the phase hologram is, by (15), (17) and (23),%

\begin{equation}
\Psi(x,z)=-i\ln\left[  \psi(x+\hbar z/2)\psi^{\ast}(x-\hbar z/2)\right]  .
\end{equation}

It will be shown next that, \emph{alternatively}, a one dimensional
convolution \emph{along} $x$ can arise in the representation for $p(x,\mu)$.

\section{Alternative case of spread in $x$}

The preceding analysis indicated that a conditional PDF $g(\mu|x)$ convolved
with the Wigner distribution is the answer for the net $p(x,\mu).$ \ Note that
that derivation started with the definition (4) of $p(x,\mu)$, whereby
variable $z$ of its spectrum $Z(x,z)$ is integrated over. \ We can instead
choose to work with a representation where the other variable $x$ of
$Z(x,z)$\ is effectively integrated over, as%

\begin{equation}
p(x,\mu)\equiv\frac{1}{2\pi}\int dkZ(k,\mu)e^{-2ikx/\hbar}.
\end{equation}
[cf. Eq. (4)]. \ Going through the analogous EPI derivation in Secs.
2.7-3.2\ now gives%

\begin{equation}
p(x,\mu)=g(x|\mu)\otimes p_{W}(x,\mu)\equiv\int dkg(x-k|\mu)p_{W}(k,\mu)
\end{equation}
[cf. Eq. (19)], where now%

\begin{equation}
p_{W}(k,\mu)=\frac{1}{\pi\hbar}\int dke^{-2ixk/\hbar}\varphi^{\ast}%
(\mu+k)\varphi(\mu-k)
\end{equation}
[cf. Eq. (1)] in terms of the momentum eigenfunction $\varphi(\mu)$ of the
system .\ 

The convolution in (26) is again one-dimensional, but now along $x$.
\ \ Effectively, this means the presence of noise of detection of
\emph{position, }rather than in momentum as in the preceding. \ The notation
$g(x|\mu)$ signifies a conditional PDF on values of $x$ in the presence of
each fixed value of momentum $\mu.$ \ As before, this is generally
signal-dependent noise, and also equates to coarse graining of coordinate
position space. \ The kernel $g(x|\mu),$ as a specifier of detection noise, is
often called the \emph{instrument function} of the measurement experiment.

We need to show next that either of the \emph{one-dimensional} convolution
answers (19) or (26) suffices to give a \emph{positive}\ $p(x,\mu)$. \ This
question seems to have not been addressed before in the voluminous literature
on the Wigner distribution.

\subsection{Positivity property of coarse graining in $\mu$}

We first treat convolution along momentum coordinate $\mu$ , then
alternatively along $x$. \ \ 

Eq. (19) in Fourier space is%

\begin{equation}
Z(x,z)=\hat{p}_{W}(x,z)|Z(x,z)|
\end{equation}
where a karat indicates the Fourier transform of the function beneath it.
\ Eqs. (3) and (20) were also used. \ By Eq. (22),%

\begin{equation}
\hat{p}_{W}(x,z)=\psi^{\ast}(x-\hbar z/2)\psi(x+\hbar z/2).
\end{equation}
So far function $|Z(x,z)|$ is arbitrary. \ Temporarily let \ %

\begin{equation}
|Z(x,z)|\equiv f_{X}(x)\exp(-\sigma^{2}z^{2}/2),~~~f_{X}(x)\geq0,~~\sigma
\text{ large,}%
\end{equation}
with $f_{X}(x)$ a positive but otherwise arbitrary function. \ Combining Eqs.
(24)-(26) and (30) give%

\begin{equation}
Z(x,z)\approx f_{X}(x)\lim_{\sigma\rightarrow\infty}\psi^{\ast}(x-\hbar
z/2)\psi(x+\hbar z/2)\exp(-\sigma^{2}z^{2}/2).
\end{equation}
Then by (3),%

\begin{equation}
p(x,\mu)=\frac{f_{X}(x)}{2\pi}\int dz\lim_{\sigma\rightarrow\infty}%
\exp(-\sigma^{2}z^{2}/2)\psi^{\ast}(x-\hbar z/2)\psi(x+\hbar z/2)\exp(-iz\mu).
\end{equation}

With $\sigma$ sufficiently large only integrand values $z\approx0$ contribute
to the integral, so that%

\begin{equation}
p(x,\mu)\approx\frac{f_{X}(x)}{2\pi}dz~\psi^{\ast}(x)\psi(x).
\end{equation}
This obeys positivity since $~f_{X}(x)\geq0$ by (30),\ since $dz>0$ and
because $\psi^{\ast}(x)\psi(x)=p(x)$ is a well-defined probability law on
position fluctuation $x$. \ Expanding out factors $\psi^{\ast}(x-\hbar z/2)
$\ and $\psi(x+\hbar z/2)$\ in (32) about the point $x$ would allow one to
extend the property of positivity for finite values of $\sigma$ as well. \ \ A
minimum necessary grain size $\sigma$ for achieving positivity for a given
state $\psi(x)$ could be found in this way. \ This has yet to be done.

\subsection{Positivity property of coarse graining in $x$}

Results for convolution along $x,$ instead, are completely analogous. \ In
place of (32), one gets%

\begin{equation}
p(x,\mu)=\frac{f_{M}(\mu)}{2\pi}\int dk\lim_{\sigma\rightarrow\infty}%
\exp(-\sigma^{2}k^{2}/2)\varphi^{\ast}(\mu+k)\varphi(\mu-k)\exp(-2ikx/\hbar)
\end{equation}
where $f_{M}(\mu)\geq0.$\ \ Again taking the $\lim\sigma\rightarrow\infty,$
only values of $k\approx0$ contribute, so that%

\begin{equation}
p(x,\mu)\approx\frac{f_{M}(\mu)}{2\pi}dk~\varphi^{\ast}(\mu)\varphi(\mu).
\end{equation}
This is again positive. \ 

\section{Amount of received information}

The first tenet of EPI is the variational principle (2), $I\ -J=extrem.$
\ This was used to form the above solution. \ The second tenet [12]-[15] is
that $I=\kappa J$ at solution, with $\kappa$ a constant on the interval
$(0,1).$\ Here, by \ (9), $J=0$. \ Therefore the data information (11) at
solution should likewise be zero. \ This is an important check on the
consistency of the theory, and we show in the Appendix that it is satisfied,
for a wide class of wave functions $\psi(x).$ \ These are those that either
have the same values at their boundaries or, if the common boundary value is
zero, approach zero at the same rate. \ The simplest example is a free
particle in a box of length $2b$, $b$ finite. \ More general classes include
wave functions having positive parity; or wave functions with periodic
boundary conditions.

\section{Discussion}

This approach has shown that the Wigner law results from the following
gedanken experiment: \ A complex hologram (3) is made from the unknown joint
law $p(x,\mu)$ for a system of state $\psi(x).$ \ A hologram is constructed
from its phase part. \ This phase hologram is illuminated by a uniform
reference beam of particles. \ One of these is randomly selected and measured
for its position as it emerges. \ For the measured position to contain an
extreme level of Fisher information about its ideal position, the joint law
$p(x,\mu)$ must obey the convolution of the Wigner law (1) with a one
dimensional PDF on either $x$ or $\mu.$ \ That is, $p(x,\mu)$ must obey
results (19), (22) and (26). \ These show a new significance for the Wigner
law: \ \emph{The Wigner distribution represents an optimal joint distribution
for purposes of conveying information about two-dimensional particle position
in a phase hologram (3) constructed out of }$p(x,\mu).$ \ 

Eqs. (19), (22)\ and (26) are also interesting in being new answers to the
problem. \ Thus, the EPI approach to the problem did not result in purely the
Wigner law, or even in the Wigner law convolved with a two-dimensional kernel
function. \ Rather, the result is the Wigner law convolved with a suitable
\emph{one dimensional} kernel. This was shown [Eqs. (33), (35)] to obey the
required property of positivity for a PDF under sufficiently coarse graining
in one dimension. \ As discussed (Sec. 1), the answer is specific to the given
measurement problem, and does not represent all measurement problems. \ Also,
it holds for a class of system states $\psi(x)$\ (see Appendix) and not all.
\ The convolution is due to noise of detection of either momentum or position
(but not necessarily both). \ This noise is the origin of coarse graining in
this problem.

It was found that, under coarse graining in either of position- or momentum
space, the law achieves positivity and therefore becomes a legitimate PDF.
\ Hence, the EPI\ result agrees with the conventional quantum view that, on
the point level $(x,\mu),$\ a joint probability law $p(x,\mu)$ on intrinsic
fluctuations does not generally exist, but that when localized averaging can
be taken, it does exist. \ However, the result differs from the conventional
view in showing that \emph{the averaging arises naturally out of the
measurement process.} \ It does not arise as merely an ad hoc appendage to the
Wigner answer. \ Also, the averaging need not be done over \emph{both}
coordinates $x$ and $\mu$. \ It is done over either \emph{the momentum}
coordinate or \emph{the position} coordinate.

Finally, the results (19), (22) obey self-consistency in obeying the
assumption $J=0$ [Eq. (9)] made in their derivation. \ Information $J$
generally represents the source information for a given problem. \ Here this
is the information that exists at the point level of $(x,\mu)$ space.\ \ With
the ansatz $J=0,$ we assumed that such information was not available. \ Either
convolution (19) or (26) bears this out, stating that no such information is
present because of the convolution (smearing) operation along either
coordinate $\mu$ or $x.$

\appendix

\section{Appendix: Systems obeying $I=0$ at solution}

By Eqs. (3), (16) and (23)%

\begin{equation}
\Psi(x,z)=-i\left[  \ln\psi(x+\alpha z)+\ln\psi^{\ast}(x-\alpha z)\right]
,~~\alpha\equiv\hbar/2.
\end{equation}
Differentiating gives%

\begin{align}
\Psi_{X}  &  =-i\left[  (\ln\psi)_{x+\alpha z}^{\prime}+(\ln\psi)_{x-\alpha
z}^{\prime}\right]  ,\\
\Psi_{Z}  &  =-i\alpha\left[  (\ln\psi)_{x+\alpha z}^{\prime}-(\ln
\psi)_{x-\alpha z}^{\prime}\right]  .\nonumber
\end{align}
Here the subscript $X$ means $\partial/\partial x$, subscript $Z$ means
$\partial/\partial z,$ and the notation $(\ln\psi)_{x\pm\alpha z}^{\prime}$
means $\left(  \partial\ln\psi(w)/\partial w\right)  $\ evaluated respectively
at $w=x\pm\alpha z$. At this point the calculation simplifies if we assume
that $\psi(x)$ is purely real. \ Full complexity is retrieved at the end.

By Eq. (11), with $\psi(x)$ purely real%

\begin{equation}
I=\int\int dxdz\left(  \Psi_{X}^{2}-\frac{1}{\alpha^{2}}\Psi_{Z}^{2}\right)
,~\
\end{equation}
where we ignore an inconsequential muliplier. \ Substituting in Eqs. (A2) and
doing the indicated squaring gives just the cross term%

\begin{equation}
I=\int\int dxdz(\ln\psi)_{x+\alpha z}^{\prime}(\ln\psi)_{x-\alpha z}^{\prime}.
\end{equation}
An irrelevent multiplier was again ignored.

It is convenient to change variables, as%

\begin{equation}
w\equiv x+\alpha z,~~v\equiv x-\alpha z.
\end{equation}
Then $x=2^{-1}(w+v),$ $z=(2\alpha)^{-1}(w-v).$ \ The Jacobian of the
transformation is then $|J(x,z/w,v)|=$ $(2\alpha)^{-1},$ so that (A4) becomes%

\begin{equation}
I=\int\int dwdv|J(x,z/w,v)|(\ln\psi(w))^{\prime}(\ln\psi(v))^{\prime}=\int\int
dwdv(\ln\psi(w))^{\prime}(\ln\psi(v))^{\prime}%
\end{equation}
after use of the particular Jacobian, and as usual ignoring a multiplicative
constant. \ The integral is actually a simple square,%

\begin{equation}
I=\left[  \int dw(\ln\psi(w))^{\prime}\right]  ^{2}.
\end{equation}
This is easily evaluated, as%

\begin{equation}
I=\left[  \int dw\frac{d}{dw}\ln\psi(w)\right]  ^{2}=\left[  \ln\psi
(b)-\ln\psi(a)\right]  ^{2}=\ln^{2}\left[  \frac{\psi(b)}{\psi(a)}\right]
\end{equation}
where $a,b$ are the boundary values of $w$. \ With the original boundaries at
$x=\pm x_{0},$ $z=\pm z_{0},$ Eqs. (A5) show that $b=x_{0}+\alpha z_{0},$
$a=-b,$ so that%

\begin{equation}
I=\lim_{x\rightarrow b}\ln^{2}\left[  \frac{\psi(x)}{\psi(-x)}\right]
,~~b=x_{0}+\alpha z_{0}.
\end{equation}
This shows that $I=0$ if $\psi(b)\rightarrow\psi(-b)$, i.e. the same value
$\psi$\ is approached as the particle approaches its boundaries. \ An example
is a system whose $\psi$ has positive parity. \ Another is where the system is
periodic, with period $2b$. \ A third is not necessarily periodic in all its
values, but has boundary values that repeat. \ A possible complication is
where the boundary values are zero, as in bound systems, since $0/0$ is
indeterminate. \ Here what is required is that $\psi$ approaches zero
symmetrically at the two boundaries of this system. \ This is obeyed by a wide
class of bound systems. \ \ Examples are states with positive parity; or, as
the simplest example, a free particle in a box of length $2b$, $b$ finite,
where $\lim_{x\rightarrow\pm a}$ $\psi(x)=$ $\lim_{x\rightarrow\pm b}\cos(n\pi
x/2b)=+0,$ $n=1,3,...$ (Here both boundary values approach zero symmetrically
from above.)

We assumed a real $\psi(x),$ for simplicity in deriving (A9). \ With $\psi(x)$
instead complex, the net $I$ turns out to be the quantity (A9) plus the same
expression in $\psi^{\ast}.$ \ The latter expression is zero under the
condition $\psi^{\ast}(b)=\psi^{\ast}(-b),$ which is the same condition as before.

In summary the condition $I=0$\ holds, not for all systems, but for a wide
class of systems.\ \ \ This is a natural consequence of the overall approach
which, as discussed (Sec. I), is specific to the given measurement problem. \ \ 

\section{}

[1] E. Wigner, Phys. Rev. \textbf{40} (1932) 749

[2] N.C. Dias, J.N. Prata, "Admissible states in quantum phase space," Ann.
Phys. \textbf{313,}110 (2004) ; also arXiv:hep-th/0402008 v1 1 Feb 2004

[3] E.B. Davies, \emph{Quantum Theory of Open Systems} (Academic, New York, 1976)

[4] J. Uffinck, "The joint measurement problem," Int. J. Theor. Phys.
\textbf{33}, 199 (1994)

[5] K. Wodkiewicz, Pys. Rev. Lett. \textbf{52}, 1064 (1984). \ We quote:\ "Is
it possible to define a realistic phase-space function that can be
\emph{recorded} [our italics] in the laboratory?"

[6] \ N.D. Cartwright, Physica \textbf{83A}, 210 (1976); L. Diosi,
arXiv:quantum-ph/0212103 v1 17 Dec 2002, 1-6

[7] A. Wax and J.K. Thomas, Optics Lett. \textbf{21}, 1427 (1996)

[8] L. McMackin, D.G. Voeltz, M.P. Fetrow, Optics Express \textbf{1}, 332 (1997)

[9] R. Thalmann and R. Dandliker, Proc. SPIE \textbf{599}, 141 (1985)

[10] P. Hariharan, Opt. Eng. \textbf{24}, 632 (1985)

[11] P.K. Rastogi, ed., \emph{Holographic interferometry}, Springer Ser. Opt.
Sci. 68 (Springer-Verlag, Berlin, 1994)

[12] B.R. Frieden and B.H. Soffer, Phys. Rev. E \textbf{52}, 2274 (1995)

[13] B.R. Frieden, \emph{Physics from Fisher Information} (Cambridge
University Press, 1998)

[14] B.R. Frieden, \emph{Science from Fisher Information} (Cambridge
University Press, 2004)

[15] B.R. Frieden and R.A. Gatenby, eds., \emph{Exploratory Data Analysis
Using Fisher Information }(Springer-Verlag, London, 2006; in press)

[16] D. Kermisch, J. Opt. Soc. Am. \textbf{60}, 15 (1970). \ We quote:\ "The
phase information thoroughly dominates the process" [of image reconstruction].
\ "Up to 78\% of the total image radiance reconstructs exactly the original
image. \ The other 22\% consists of [self] convolutions of the original image,
and degrades the recontructed image.

[17] S. Furhapter, A. Jesacher, S. Bernet and M.\ Ritsch-Marte, Opt. Express
\textbf{13}, 689 (2005)

[18] S.A. Benton, "Holographic Displays: 1975-1980," Opt. Eng. \textbf{19},
686 (1980)

[19] A. Marquez, J. Campos and M.J. Yzuel, Opt. Eng. \textbf{39}, 1612 (2000)

[20] Use of an imaginary coordinate is by so-called \emph{Wick rotation}.
\ The imaginary time coordinate was invented by H. Minkowski for defining a
relativistic four-space. \ A recent use of the Wick rotation is H. Gies and K.
Klingmuller, Phys. Rev. D \textbf{72}, 065001 (2005)

[21] V.I. Tatarskii, Sov. Phys. Usp. \textbf{26}, 311 (1983)

[22] B.R. Frieden, \emph{Probability, Statistical Optics and Data Testing, 3rd
ed.} (Springer-Verlag, Berlin, 2001)

\bigskip
\end{document}